\documentclass[aps,twocolumn,amsfonts,amssymb,amsmath,showpacs,groupedaddress]{revtex4}
\usepackage[dvips]{graphicx}

\newtheorem{theorem}{Theorem}

\begin{document}
\title{Quantum Discord and Geometry for a Class of Two-qubit States}

\author{Bo Li}
\affiliation{School of Mathematical Sciences, Capital Normal
University, Beijing 100048, China}
\affiliation{Department of Mathematics and Computer, Shangrao Normal University,
 Shangrao 334001, China}
\author{Zhi-Xi Wang}
\affiliation{School of Mathematical Sciences, Capital Normal
University, Beijing 100048, China}
\author{Shao-Ming Fei}
\affiliation{School of Mathematical Sciences, Capital Normal
University, Beijing 100048, China} \affiliation{Max-Planck-Institute
for Mathematics in the Sciences, 04103 Leipzig, Germany}

\begin{abstract}
We study the level surfaces of quantum discord for a class of two-qubit states
with parallel nonzero Bloch vectors. The dynamic behavior of quantum discord under decoherence is investigated.
It is shown that a class of $X$ states has sudden transition between classical
and quantum correlations under decoherence. Our results include the ones in
[Phys. Rev. Lett. 105. 150501] as a special case and show new pictures and structures of quantum discord.
\end{abstract}
\pacs{03.67.Mn, 03.65.Ud,  03.65.Yz}
\maketitle
\section{\bf Introduction}

The quantum entanglement acts as the most important resources
in quantum information \cite{horodecki1,nielsen}. However, entanglement is not
the only correlation that is useful for quantum information processing. Recently, it is found
that many tasks, e.g. quantum nonlocality without entanglement \cite{horodecki1,bennett,niset},
can be carried out with quantum correlations other than entanglement.
It has been shown both theoretically and experimentally \cite{datta,lanyon}
that some separable states may speed up certain tasks over their classical counterparts.

One kind of nonlocal correlation called quantum discord, as introduced by
Oliver and Zurek \cite{ollivier}, has received much attention recently \cite{ollivier,bylicka,werlang,sarandy,ferraro,fanchini,dakic,modi,luo1,luo,lang,ali,Mazzola,Maziero}.
The idea is to measure the discrepancy between two natural yet different quantum analogs of
the classical mutual information. Let $\rho^{AB}$ denote the density operator of
a composite bipartite system AB,  and $\rho^{A(B)}=Tr_{B(A)}(\rho^{AB})$
the reduced density operator of the partition B(A). The
quantum mutual information is defined by
\begin{eqnarray}
\mathcal{I}(\rho^{AB})=S(\rho^A)+S(\rho^B)-S(\rho^{AB}),
\end{eqnarray}
where $S(\rho)=-Tr(\rho\log_2\rho)$ is the Von Neuman entropy.
It was shown that quantum mutual information is the information-theoretic
measure of the total correlation in a bipartite quantum state. In order to
determine quantum discord \cite{ollivier,luo}, Ollivier and Zurek use a
measurement-based conditional density operator to generalize the classical
mutual information. Let ${B_k}$ be a set of one-dimensional project measurement performed
on subsystem $B$, the conditional density operator $\rho_k$ associated with the
measurement result $k$ is
 \begin{eqnarray}
\rho_k=\frac{1}{p_k}(I\otimes B_k)\rho (I\otimes B_k),
\end{eqnarray}
where $p_k=tr(I\otimes B_k)\rho (I\otimes B_k)$, $I$ is the identity operator on the subsystem $A$.
With this conditional density operator, the quantum conditional entropy with respect to this measurement is
defined by
\begin{eqnarray}
S(\rho|\{B_k\}):=\sum_{k}p_kS(\rho_k),
\end{eqnarray}
and the associated quantum mutual information is given by
\begin{eqnarray}
\mathcal{I}(\rho|\{B_k \}):=S(\rho^A)-S(\rho|\{B_k\}).
\end{eqnarray}
Classical correlation is defined as the superior of $\mathcal{I}(\rho|\{B_k\})$ over
all possible Von Neumann measurement ${B_k}$,
\begin{eqnarray}
\mathcal{C}(\rho):=\sup_{\{B_k\}}I(\rho|\{B_k\}).
\end{eqnarray}
Quantum discord is then given by the difference of mutual information $\mathcal{I}(\rho)$ and
the classical correlation $\mathcal{C}(\rho)$,
\begin{eqnarray}
\mathcal{Q}(\rho):=\mathcal{I}(\rho)-\mathcal{C}(\rho).
\end{eqnarray}

The analytical expressions for classical correlation and quantum discord are only
available for two-qubit Bell diagonal state and a seven-parameter family of two-qubit $X$ states \cite{luo,ali} till now.
For the two-qubit Bell-diagonal state:
\begin{eqnarray}
\rho=\frac{1}{4}(I\otimes I+\sum_{i=1}^3c_i\sigma_i\otimes\sigma_i),\label{bellstate}
\end{eqnarray}
the classical correlation is given by
\begin{eqnarray}
\mathcal{C}(\rho)=\frac{1-c}{2}\log_2(1-c)+\frac{1+c}{2}\log_2(1+c),
\end{eqnarray}
where $c=\max\{|c_1|,|c_2|,|c_3|\}$. The quantum discord is given by
\begin{eqnarray}
\mathcal{Q}(\rho)&=&\frac{1-c_1-c_2-c_3}{4}\log_2(1-c_1-c_2-c_3)\nonumber\\
& &+\frac{1-c_1+c_2+c_3}{4}\log_2(1-c_1+c_2+c_3)\nonumber\\
& &+\frac{1+c_1-c_2+c_3}{4}\log_2(1+c_1-c_2+c_3)\nonumber\\
& &+\frac{1+c_1+c_2-c_3}{4}\log_2(1+c_1+c_2-c_3)\nonumber\\
& &-\frac{1-c}{2}\log_2(1-c)-\frac{1+c}{2}\log_2(1+c).
\end{eqnarray}

The geometry of Bell-diagonal states was first introduced by Horodecki \cite{horodecki2}.
From the positivity of the spectral of a Bell-diagonal state $\rho$ in Eq.(\ref{bellstate}),
one can see that $\rho$ belongs to a tetrahedron ${\cal T}$ with vertices $v_1=(1,-1,1)$, $v_2=(-1,1,1)$, $v_3=(1,1,-1)$,
and $v_4=(-1,-1,-1)$ in the correlation vector space. Similarly, from the positivity of the
partial transpose of $\rho$, it has been shown that
that the separable states belong to the octahedron ${\cal O}$ with
vertices $O_1^\pm=(\pm1,0,0)$, $O_2^\pm=(0,\pm1,0)$ and $O_3^\pm=(0, 0 ,\pm 1)$ \cite{horodecki2,kim,lang}.

Very recently, Matthias D. Lang and Carlton M. Caves \cite{lang} depicted
the level surfaces of entanglement and quantum discord for Bell-diagonal states,
they discovered that the picture and the structure of the quantum entanglement and the quantum discord
are very different. There doesn't exist simple relations between them.

In this article, we study the quantum discord for a class of $X$ states that the Bloch vectors are $z$ directional,  which including
Bell-diagonal states as a special case.
We study the level surfaces of quantum discord and dynamic behavior of quantum discord under decoherence.
It is demonstrated that the surfaces of constant discord shrinks along
with the geometrical deformation of ${\cal T}$ in Ref.\cite{kim}. Moreover
we find that there is a class of $X$ states for which the quantum discord is not destroyed by
decoherence in a finite time interval.

We calculate different kinds of correlation such as entanglement, classical correlation and quantum discord
for the state we concerned in sec. \ref{surface}. We depict the level surface of constant discord in four
different situations.  In sec. \ref{dynamics}, we discuss the dynamics of quantum discord
and show that the quantum discord of a certain class of $X$ states does not decay under decoherence.
A brief conclusion is given in sec. \ref{discuss}.

\section{\bf Geometrical depiction of ${\cal C}$  and ${\cal D}$  }\label{surface}
Under appropriate local unitary transformations, any two-qubit state $\rho$ can be written as:
\begin{eqnarray}
\rho=\frac{1}{4}[I\otimes I+\textbf{r}\cdot\sigma\otimes I+I\otimes\textbf{s}\cdot\sigma+\sum_{i=1}^3c_i\sigma_i\otimes\sigma_i],
\label{twoqubitdiaostate}
\end{eqnarray}
where \textbf{r} and \textbf{s} are Bloch vectors and $\{\sigma_i\}_{i=1}^3$ are
the standard Pauli matrices. When \textbf{r}=\textbf{s}=\textbf{0},
$\rho$ reduces to the two-qubit Bell-diagonal states.
In the following, we assume that the Bloch vectors are $z$ directional, that is, $\textbf{r}=(0,0,r)$, $\textbf{s}=(0,0,s)$.
One can also change them to be $x$ or $y$ directional via
an appropriate local unitary transformation without losing its diagonal property of the correlation term  \cite{kim}.
In this case the arbitrary state $\rho$ defined in Eq.(\ref{twoqubitdiaostate}) has the form
\begin{widetext}
\begin{eqnarray}
\rho = \frac{1}{4} \left(
\begin{array}{cccc}
1+r+s+c_3
& 0 & 0 & c_1 -c_2 \\
0 & 1+r-s-c_3 & c_1+c_2 & 0 \\
0 & c_1 +c_2 & 1-r+s-c_3
& 0 \\
c_1 -c_2 & 0 & 0 & 1-r-s+c_3
\end{array}
\right) \,.
\label{Eq:AXstate}
\end{eqnarray}
\end{widetext}

The entanglement of formation \cite{wootters} is a monotonically increasing
function of the Wootter's concurrence. While the concurrence can be
calculated in terms of the eigenvalues of $\rho\widetilde{\rho}$, where $\widetilde{\rho}=\sigma_y\otimes \sigma_y\rho^*\sigma_y\otimes \sigma_y$.
For the state Eq.(\ref{Eq:AXstate}), the eigenvalues of $\rho\widetilde{\rho}$ are
\begin{eqnarray}
\lambda_1&=&\frac{1}{16}(c_1-c_2-\sqrt{(1+c_3)^2-(r+s)^2})^2\nonumber\\
&=&\frac{1}{16}(c_1-c_2-\sqrt{(1+r+s+c_3)(1-r-s+c_3)})^2\nonumber,
\end{eqnarray}
\begin{eqnarray}
\lambda_2&=&\frac{1}{16}(c_1-c_2+\sqrt{(1+c_3)^2-(r+s)^2})^2\nonumber\\
&=&\frac{1}{16}(c_1-c_2+\sqrt{(1+r+s+c_3)(1-r-s+c_3)})^2\nonumber,
\end{eqnarray}
\begin{eqnarray}
\lambda_3&=&\frac{1}{16}(c_1+c_2-\sqrt{(1-c_3)^2-(r-s)^2})^2\nonumber\\
&=&\frac{1}{16}(c_1+c_2-\sqrt{(1+r-s-c_3)(1-r+s-c_3)})^2\nonumber,
\end{eqnarray}

\begin{eqnarray}
\lambda_4&=&\frac{1}{16}(c_1+c_2+\sqrt{(1-c_3)^2-(r-s)^2})^2\nonumber\\
&=&\frac{1}{16}(c_1+c_2+\sqrt{(1+r-s-c_3)(1-r+s-c_3)})^2\nonumber.
\end{eqnarray}
The concurrence is given by
\begin{widetext}
\begin{eqnarray}
C(\rho)=\max\{2\max\{\sqrt{\lambda_1},\sqrt{\lambda_2},\sqrt{\lambda_3},\sqrt{\lambda_4}\}
-\sqrt{\lambda_1}-\sqrt{\lambda_2}-\sqrt{\lambda_3}-\sqrt{\lambda_4},0 \}.
\label{twoqubitconcurrence}
\end{eqnarray}
\end{widetext}

If one fixes the parameters $r$ and $s$, the above states and their concurrence are a three parameters set, with
the Bell-diagonal states belonging to the set with $r=s=0$. The geometry of
such set with nonzero Bloch vectors has been considered by Hungsoo Kim \textit{et al.} recently \cite{kim}.
The geometrical deformation of the octahedron ${\cal T}$ for the
set of  Bell-diagonal states and the octahedron ${\cal O}$ for the separable Bell-diagonal states has been depicted.
The deformation of ${\cal O}$ can also be obtained from the region where $C(\rho)=0$ in Eq.(\ref{twoqubitconcurrence}),
as the concurrence of separable state must be zero. The level surfaces of concurrence or entanglement can
be plotted correspondingly.

As $\rho$ in Eq.(\ref{Eq:AXstate}) is a two-qubit X state, the discord
can be calculated in a way presented in \cite{ali}.
The eigenvalues of $\rho$ in Eq.(\ref{Eq:AXstate}) is given by
$$
\begin{array}{l}
u_\pm=\frac{1}{4}[1-c_3\pm\sqrt{(r-s)^2+(c_1+c_2)^2} ],\\[2mm]
v_\pm=\frac{1}{4}[1+c_3\pm\sqrt{(r+s)^2+(c_1-c_2)^2} ].
\end{array}
$$
For convenience, we define $f(t)=-\frac{1-t}{2}\log_2(1-t)-\frac{1+t}{2}\log_2(1+t)$.
$f(t)$ is a monotonically decreasing function for $0\leq t\leq 1$.
The quantum mutual information is given by
\begin{eqnarray}
\mathcal{I}(\rho)&=&S(\rho^A)+S(\rho^B)+u_+\log_2u_+\nonumber\\
&+&u_-\log_2u_-+v_+\log_2v_++v_-\log_2v_-,
\label{mutualinformation}
\end{eqnarray}
where $S(\rho^A)$ and $S(\rho^B)$ are given by $S(\rho^A)=1+f(r)$, $S(\rho^B)=1+f(s)$.

We evaluate next the classical correlation $\mathcal{C}(\rho)$. The Von Neumann measurement
for subsystem $B$ can be written as $B_i=V\prod_iV^+$, $i=0,1$, where $\prod_i=|i\rangle\langle i|$
is the projector associated with the subsystem $B$ and $V=tI+i\overrightarrow{y}\cdot\overrightarrow{\sigma}\in SU(2)$,
$t, y_1, y_2, y_3\in R$ and $t^2+y_1^2+y_2^2+y_3^2=1$.
After the measurement, we have the ensemble $\{\rho_i, p_i\}.$
The classical correlation is therefore given by
\begin{eqnarray}
\mathcal{C}(\rho) &=& \sup_{\{B_i\}} \, \mathcal{I} (\rho|\{B_i\}) \nonumber\\
   &=& S(\rho^A)- \min_{\{B_i\}} S (\rho|\{B_i\}),        \label{Eq:CC}
\end{eqnarray}
where
\begin{eqnarray}
S(\rho|\{B_i\})=p_0S(\rho_0)+p_1S(\rho_1).
\label{conditionalentropy}
\end{eqnarray}
By a the parameter transformation
$$
\begin{array}{c}
m=(ty_1+y_2y_3)^2, ~n=(ty_2-y_1y_3)(ty_1+y_2y_3),\\[2mm]
k=t^2+y_3^2,~ l=y_1^2+y_2^2,
\end{array}
$$
which satisfies $m^2+n^2=klm$, $k+l=1, k\in[0,1]$, $m\in[0,\frac{1}{4}]$ and $n\in[-\frac{1}{8},\frac{1}{8}]$,
according to \cite{ali} we observe that the minimum of Eq.(\ref{conditionalentropy}) can
only be obtained in the following cases:

(1) $k=1$, $l=0$, $m=n=0.$ For state (\ref{Eq:AXstate}),
Eqs.(14-17) in Ref.\cite{ali} turn out to be $p_0=\frac{1+s}{2},$ $p_1=\frac{1-s}{2},$
$\theta=\mid \frac{r+c_3}{1+s}\mid$, $\theta'=\mid \frac{r-c_3}{1-s}\mid$,
$v_\pm (\rho_0)=\frac{1\pm\theta}{2}$, $\omega_\pm (\rho_1)=\frac{1\pm\theta'}{2}$.
Thus,
\begin{eqnarray}
S_1&=& S(\rho|\{B_i\})=p_0S(\rho_0)+p_1S(\rho_1) \nonumber\\
   &=& -\frac{1+r+s+c_3}{4}\log_2\frac{1+r+s+c_3}{2(1+s)}\nonumber\\
    &-&\frac{1-r+s-c_3}{4}\log_2\frac{1-r+s-c_3}{2(1+s)}\nonumber\\
     &-&\frac{1+r-s-c_3}{4}\log_2\frac{1+r-s-c_3}{2(1-s)}\nonumber\\
      &-&\frac{1-r-s+c_3}{4}\log_2\frac{1-r-s+c_3}{2(1-s)}.      \label{Eq:S1}
\end{eqnarray}

(2) $k=0$, $l=1$, $m=n=0$. It is easy to find that the minimum is the same as $S_1$.

(3) $k=l=\frac{1}{2}$. In this case, we have
$$\theta=\theta'=\sqrt{r^2+c_1^2-4m(c_1^2-c_2^2)},$$
here $\theta,\, \theta'$ are defined by Eqs. (16,17) in \cite{ali},
$S(\rho_0)=S(\rho_1),$ which is a monotonically function of $m$. Therefore the minimum is
obtained at $m=0$ or $m=\frac{1}{4}$. We have either $\theta=\theta'=\sqrt{r^2+c_1^2}$
or $\theta=\theta'=\sqrt{r^2+c_2^2}$. The quantum conditional entropy is given by
\begin{eqnarray}
S_2=1+f(\sqrt{r^2+c_1^2}),      \label{Eq:S2}
\end{eqnarray}
\begin{eqnarray}
S_3=1+f(\sqrt{r^2+c_2^2}).      \label{Eq:S3}
\end{eqnarray}
Therefore, we have
\begin{theorem}
For any state $\rho$ of the form Eq.(\ref{Eq:AXstate}), the classical correlation of $\rho$ is given by
\begin{eqnarray}
\label{proposition1}
\mathcal{C}(\rho)= S(\rho^A)- \min\{S_1, S_2, S_3\},
\end{eqnarray}
where $S_1, S_2, S_3$ are defined by Eqs.(\ref{Eq:S1}), (\ref{Eq:S2}), (\ref{Eq:S3}) respectively.
The quantum discord is given by
\begin{eqnarray}
\mathcal{Q}(\rho)=\mathcal{I}(\rho)-\mathcal{C}(\rho),
\end{eqnarray}
with $\mathcal{I}(\rho)$ given by (\ref{mutualinformation}).
\end{theorem}

In Fig.1 we plot the level surface of discord when (a) $r=s=0.3$, $\mathcal{Q}(\rho)=0.03$;
(b) $r=s=0.5$, $\mathcal{Q}(\rho)=0.03$; (c) $r=s=0.3$, $\mathcal{Q}(\rho)=0.15$; (d) $r=s=0.5$, $\mathcal{Q}(\rho)=0.15$.
From Fig.1 one can see that the level surface of discord has a great change from the case $r=s=0$ studied in Ref. \cite{lang}.
The surface shrinks with the effect of $r$ and $s$ and the shrinking rate becomes larger with the increasing $|r|$ and $|s|$.
What is more, when the discord is small (such as $\mathcal{Q}(\rho)=0.03$), the horizontal ``tubes'' are closed! see Fig (a).
For larger $r$ and $s$, the picture is moved up the plane $c_3=0$, see Fig. (b).
For larger discord and small $r$ and $s$, Fig. (c), the figure is similar to the ones in case of $r=s=0$.
But for larger $r$ and $s$, Fig. (d), the figure is moved up again and changes dramatically also.

\begin{figure}[h]
\begin{center}
\caption{Surfaces of constant discord: (a) $r=s=0.3$, $\mathcal{Q}(\rho)=0.03$;
(b) $r=s=0.5$, $\mathcal{Q}(\rho)=0.03$; (c) $r=s=0.3$, $\mathcal{Q}(\rho)=0.15$;
(d) $r=s=0.5$, $\mathcal{Q}(\rho)=0.15$.}
\end{center}
\label{Fig:4}
\end{figure}

\section{\bf Dynamics of quantum discord under local nondissipative channels}\label{dynamics}
It has been recently discovered that for some Bell-diagonal states, their quantum discord are invariant under
some decoherence for a finite time interval \cite{Mazzola}. An inetresting question is if such phenomena exits in other systems. In the following
we consider that the state $\rho$ in Eq.(\ref{Eq:AXstate}) undergoes the phase flip channel \cite{Maziero}, with the Kraus operators
$\Gamma_0^{(A)}=$ diag$(\sqrt{1-p/2},\sqrt{1-p/2})\otimes I$, $\Gamma_1^{(A)}=$ diag$(\sqrt{p/2},-\sqrt{p/2})\otimes I$,
$\Gamma_0^{(B)}= I \otimes$ diag$(\sqrt{1-p/2},\sqrt{1-p/2}) $, $\Gamma_1^{(B)}= I \otimes$ diag$(\sqrt{p/2},-\sqrt{p/2}) $, where $p=1-\exp(-\gamma t)$, $\gamma$ is
the phase damping rate \cite{Maziero,yu}.

Let $\varepsilon(\cdot)$ represent the operator of decoherence. Then under the phase flip channel, we have
\begin{eqnarray}
\varepsilon(\rho)&=& \frac{1}{4}(I\otimes I+r\sigma_3\otimes I+I\otimes s \sigma_3+(1-p)^2c_1\sigma_1\otimes\sigma_1\nonumber\\
    &+&(1-p)^2c_2\sigma_2\otimes\sigma_2+c_3\sigma_3\otimes\sigma_3).      \label{Eq:epsilonrho}
\end{eqnarray}

Noting that $r,\,s,\,c_3$ are independent of time, we consider the case that
\begin{eqnarray}
c_2=-c_3c_1,~ s=c_3r,~ -1\leq c_3\leq 1,~ -1\leq r\leq 1. \label{condition}
\end{eqnarray}
Then the eigenvalues of $\varepsilon(\rho)$  are given by
$$
\begin{array}{l}
u_\pm=\frac{1-c_3}{4}(1\pm \sqrt{r^2+(1-p)^4c_1^2}),\\[2mm]
v_\pm=\frac{1+c_3}{4}(1\pm \sqrt{r^2+(1-p)^4c_1^2}).
\end{array}
$$
From (\ref{mutualinformation}) we have the quantum mutual information
\begin{eqnarray}
\mathcal{I}(\varepsilon(\rho))&=&f(r)+f(c_3r)-f(c_3)-f(\sqrt{r^2+(1-p)^4c_1^2}).\nonumber\\
\label{emutualinformation}
\end{eqnarray}

To calculate the classical correlation, we need to determine $S_1$, $S_2$ and $S_3$ defined by
(\ref{Eq:S1}), (\ref{Eq:S2}) and (\ref{Eq:S3}) respectively, which are given by
\begin{eqnarray}
S_1(p)=1+f(r)+f(c_3)-f(c_3r),
\label{s11}
\end{eqnarray}
\begin{eqnarray}
S_2(p)=1+f(\sqrt{r^2+(1-p)^4c_1^2}),
\label{s22}
\end{eqnarray}
\begin{eqnarray}
S_3(p)=1+f(\sqrt{r^2+(1-p)^4c_2^2}).
\label{s22}
\end{eqnarray}
From the condition (\ref{condition}), we have $S_3(p)\geq S_2(p)$ for any $p$, while $S_2(p)$ increases under decoherence,
and $S_1(p)$ is constant under decoherence. If we select appropriate $r,\, c_1,\, c_3$ then the
initial state $S_2(0) < S_1(0)$. On the other hand, since $f(c_3)\leq f(c_3r)$, we always have
$S_2(1) \geq S_1(1)$ .
Therefore there exist $0\leq p_0\leq 1$ such that $min\{S_1,S_2,S_3\}=S_2$
for $0\leq p\leq p_0,$  and $min\{S_1,S_2,S_3\}=S_1$ for $ p_0\leq p\leq 1$.
In this case $\mathcal{Q}(\varepsilon(\rho))$ monotonically decreases to zero.

When $min\{S_1,S_2,S_3\}=S_2,$ we have
\begin{eqnarray}
\mathcal{Q}(\varepsilon(\rho))&=&\mathcal{I}(\varepsilon(\rho))-\mathcal{C}(\varepsilon(\rho))\nonumber\\
&=&f(c_3r)-f(c_3),
\end{eqnarray}
$\mathcal{Q}(\varepsilon(\rho))$ is constant under decoherence during the time interval
such that the condition $min\{S_1,S_2,S_3\}=S_2$ is satisfied.

As an example, for $r=s=0$, $c_1=1$, $-1\leq c_2=-c_3\leq1$, we have that $S_1(0)=1+f(c_3)$,
$S_2(0)=1+f(1)<S_1(0)$. Therefore the state has constant discord
under decoherence, which recovers the results in \cite{lang,Mazzola}.
For an example with nonzero $r$ and $s$,
we set $r=\frac{3}{10}$, $s=\frac{3}{20}$, $c_1^2=\frac{4}{5},$ $c_2=-\frac{c_1}{2},$
$ c_3=\frac{1}{2}$. It is direct to verify that $S_1(0)=0.762,$ $S_2(0)=0.186$.
Therefore we have $min\{S_1,S_2,S_3\}=S_2$ and the state has a constant discord. The dynamic behavior of correlation of the state  under
the phase flip channel  is depicted  in Fig.2. We find that
the concurrence $C$ is greater than the quantum discord $\mathcal{Q}$ for $0\leq p\leq 0.217$.
A sudden transition of classical and quantum correlation happens at $p=0.274$, and a
sudden death of entanglement \cite{yu1} appears at $p=0.4$.
Moreover, different from the case of zero $r$ and $s$ in \cite{Mazzola}, where the entanglement disappears before the sudden
transition of classical and quantum correlation, here one sees that the concurrence keeps non-zero
after the transition. Therefore for these states the entanglement is more robust against the decoherence
than the discord.

\begin{figure}[h]
\scalebox{2.0}{\includegraphics[width=3.25cm]{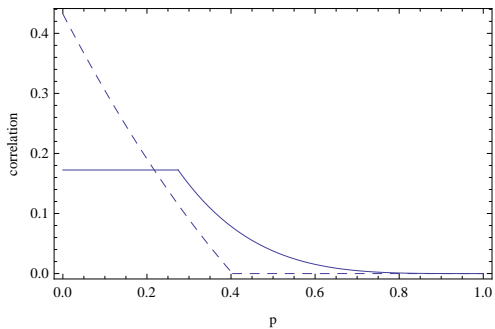}}
\caption{Concurrence(dashed line) and quantum discord(solid line) under phase flip channel for $r=\frac{3}{10}$,
$s=\frac{3}{20}$, $c_1^2=\frac{4}{5},$ $c_2=-\frac{c_1}{2}$ and $c_3=\frac{1}{2}$.}
\label{transition}
\end{figure}

\section{\bf summary}\label{discuss}
We have studied the correlation for a class of $X$ states.
The level surfaces of quantum discord have been depicted. For $r=s=0$ our results reduce to
the ones for Bell-diagonal states. For nonzero $r$ and $s$, it has been shown that the level surfaces of quantum discord
may have quite different geometry and topology. While the quantum discord
could still keep constant under decoherence in certain time interval for some initial states,
the order of sudden transition of classical and quantum correlation and the sudden death of entanglement
can be exchanged.

\bigskip
\noindent {\bf Acknowledgments} We thank Matthias D. Lang for very helpful discussions.
This work is supported by the NSFC10875081, NSFC10871227, KZ200810028013 and PHR201007107 and NSFBJ1092008.

\end{document}